\documentclass[prb,floatfix,twocolumn,showpacs,aps]{revtex4}%
\usepackage{graphicx}
\usepackage{dcolumn}
\usepackage{amsmath}
\usepackage{natbib}
\usepackage{amsfonts}
\usepackage{amsmath}
\usepackage{amssymb}
\usepackage{graphicx}%
\begin{document}
\title{Novel Pressure Induced Structural Phase Transition in AgSbTe$_{2}$ }
\author{Ravhi S. Kumar$^{a}$, Andrew L. Cornelius$^{a}$, Eunja Kim$^{a}$, Yongrong
Shen$^{a}$, S. Yoneda$^{b}$, Chenfeng Chen$^{a}$ and Malcolm F. Nicol$^{a}$}
\affiliation{$^{a}$HiPSEC and Department of Physics, University of Nevada Las Vegas, Las
Vegas, Nevada 89154, USA}
\affiliation{$^{b}$Kanagawa University, Kanagawa, Japan}

\begin{abstract}
We report a novel high pressure structural sequence for the functionally
graded thermoelectric, narrow band gap semiconductor AgSbTe$_{2}$, using angle
dispersive x-ray diffraction in a diamond anvil cell with synchrotron
radiation at room temperature. The compound undergoes a B1 to B2 transition;
the transition proceeds through an intermediate amorphous phase found between
17-26 GPa that is quenchable down to ambient conditions. The pressure induced
structural transition observed in this compound is the first of its type
reported in this ternary cubic family, and it is new for the B1-B2 transition
pathway reported to date. Density Functional Theory (DFT) calculations
performed for the B1 and B2 phases are in good agreement with the experimental results.

\end{abstract}
\date{09/30/2004}
\maketitle

Ternary semiconductors with a general formula ABX$_{2}$ (A= Cu,Ag ;
B=In,Ga,Sb; \ X= S,Se,Te) with chalcopyrite or rock salt structure are widely
used for important technical and device applications.\ These compounds are
found to be excellent candidates for the fabrication of optical frequency
converters in solid state laser systems, photovoltaic devices and development
of solar cells.\cite{Shay75,Gaber94,Soliman98} While the Cu based
chalcopyrites are mainly studied in connection with their photovoltaic
applications, the Ag compounds find importance in thermoelectric, optical
phase change and frequency conversion
applications.\cite{Feinleib71,Detemple03} Cubic AgSbTe$_{2}$ compounds with Pb
doping have been recently shown to be excellent thermoelectric materials with
high figures of merit, and are considered promising candidates for future
energy production from heat sources.\cite{Hsu03} AgSbTe$_{2}$ with In and V
doping undergo, rapid crystalline - amorphous phase transitions on local
melting; a property that is used widely to write and rewrite Compact Disks
(CD) and Digital Versatile Disks (DVD).\cite{Tominaga97,Njoroge01} In
comparison with the classical GeSbTe phase change memory alloy, AgVInSbTe is
reported to have better erasability and cyclablity in memory switching
.\cite{Tominaga97,Njoroge01,Kalb03,Nobukini95} The phase changes in these
materials are temperature driven, and the previous studies focus mainly on
doped thin films of AgSbTe$_{2}$. To our knowledge, no detailed structural
reports are available for the host compound AgSbTe$_{2}$, under different
temperature or pressure conditions.

AgSbTe$_{2}$ is isostructural to the rocksalt type II-VI chalcogenides. As the
rock salt structure of most of the II-VI compounds is sensitive to pressure,
changing from B1 to B2,\cite{Leger83,Grzybowski83} one may expect AgSbTe$_{2}$
to display a similar crystallographic transition under pressure. The aim of
the present study is to investigate AgSbTe$_{2}$ under pressure to explore the
structural similarities with its binary analogues. In this paper, we present
experimental evidence, for the first time, of a pressure induced structural
transition from B1 (crystalline) to B2 (crystalline) with an intermediate
amorphous state. This is a new pathway for this structural sequence and is
unique in the ABX$_{2}$ family. This finding may open up new directions in the
search for similar compounds of scientific and technological importance. Also,
it forces us to develop theoretical models that can explain the pressure
induced amorphization.

AgSbTe$_{2}$ has been prepared by a conventional solid state reaction method.
Stoichometric amounts of starting material were melted in a Bridgman furnace
at 873 K for 24 hours, then cooled to ambient conditions. The reacted mixture
was checked with x-ray diffraction (XRD) and found to be single phase. In each
high pressure run, a small piece of sample cut from a pellet along with a ruby
chip was loaded with silicone oil in a stainless steel gasket having a $180$
$\mu$m\ hole between $500$ $\mu$m culet diameter diamonds in a diamond anvil
cell. XRD experiments were performed using a rotating anode x-ray generator
(Rigaku) operating with an Mo target for run 1, and synchrotron x-rays with
beam size $20\times20$ $\mu$m at HPCAT, Sector 16 IDB at the Advanced Photon
Source for runs 2 and 3. The XRD patterns were collected using a MAR imaging
plate ($300\times300$ mm$^{2}$) camera with $100\times100$ $\mu$m$^{2}$ pixel
dimension for 10-20 s. The images were integrated using FIT2D
programme,\cite{Hammersley96} and structural refinements were carried out by
Rietveld method with RIETICA (LHPM) software package.\cite{Howard98} The
pressure at the sample site has been estimated using the standard ruby
fluorescence technique and the newly proposed ruby scale of
Holzapfel.\cite{Holzapfel03}

At ambient conditions, AgSbTe$_{2}$ crystallizes in the rocksalt structure
(\textit{Fm3m} symmetry); the metal atoms located at Na sites at $(0,0,0)$ and
the Te atoms\ occupy the Cl sites at $(0.5,0.5,0.5)$. The diffraction peaks at
ambient pressure can clearly be indexed to an fcc lattice with cell parameter
$a=$ $6.0780(1)$ \AA .\cite{Geller59} Upon compression, the diffraction lines
remain sharp and systematically shift with pressure as shown in Fig. 1. From
the refinements, $a$ decreases at a rate of 0.024 \AA /GPa, while the the
Ag/Sb-Te distance decreases at 0.0118 \AA /GPa. The small variation of the
Ag/Sb-Te distance relative to $a$ indicates the rigid bonding of the
metal-chalcogen atoms. The rocksalt phase is stable up to 15 GPa similar to
the behavior of the isostructural compound AgSbSe$_{2}$ under
pressure.\cite{Kumar99} At 16 GPa, we have noticed splitting of (200) and
(420) peaks (inset of Fig. 1(b)) with broadening of other diffraction lines.
These features clearly indicate the onset of a structural transformation. To
our surprise, at 17 GPa, we observed a sudden drop in the intensity of all
diffraction lines, leaving a halo peak located ar ound $2\theta\backsim
9.5^{\circ}$ with a significant broad background. The features observed in the
XRD pattern at this pressure resembled the spectra typical for amorphous
compounds.\ The amorphous phase persists up to 25 GPa as shown in Fig. 1(c).%
%TCIMACRO{\FRAME{ftbpFU}{1.7858in}{4.2065in}{0pt}{\Qcb{Representative x-ray
%diffraction patterns with increasing pressure. The inset in (b) shows the
%splitting observed in the (200) line before amorphization. The Rietveld
%refinement for the \textit{Cmcm} phase is also shown. The refinement residual
%is R$_{wp}$ = 1.473. Diffraction patterns at 24 GPa with back ground (i) and
%after back ground substraction (ii) are shown in (c). The Rietveld plot for
%the x-ray diffraction pattern obtained at 27 GPa (d) shows the B2 structure.}%
%}{\Qlb{xrd}}{kumar1.eps}{\special{ language "Scientific Word";
%type "GRAPHIC";  maintain-aspect-ratio TRUE;  display "USEDEF";
%valid_file "F";  width 1.7858in;  height 4.2065in;  depth 0pt;
%original-width 1.7487in;  original-height 4.1563in;  cropleft "0";
%croptop "1";  cropright "1";  cropbottom "0";
%filename 'kumar1.eps';file-properties "XNPEU";}}}%
%BeginExpansion
\begin{figure}
[ptb]
\begin{center}
\includegraphics[
height=4.2065in,
width=1.7858in
]%
{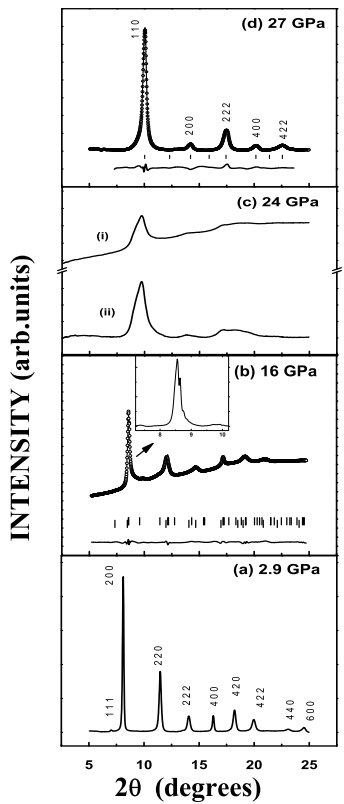}%
\caption{Representative x-ray diffraction patterns with increasing pressure.
The inset in (b) shows the splitting observed in the (200) line before
amorphization. The Rietveld refinement for the \textit{Cmcm} phase is also
shown. The refinement residual is R$_{wp}$ = 1.473. Diffraction patterns at 24
GPa with back ground (i) and after back ground substraction (ii) are shown in
(c). The Rietveld plot for the x-ray diffraction pattern obtained at 27 GPa
(d) shows the B2 structure.}%
\label{xrd}%
\end{center}
\end{figure}
%EndExpansion

The current models explaining the reconstructive phase changes of B1 structure
are based upon either the Buergers mechanism (i.e. the strain model) or the
one emphasized by Watanabe \textit{et al}. based on displacive
transitions.\cite{Sims98,Stokes02,Sowa00,Zhang03} Recently Toledano \textit{et
al}. studied different sequences for the pressure driven transitions from the
B1 phase due to the coupling of the tensile and shear
strains,\cite{Toledano03} and they suggest\ possible intermediate orthorhombic
B33 (\textit{Cmcm}) and B16 (\textit{Pbnm}) structures. Previous high pressure
structural studies on II-VI chalcogenides have shown transformation of NaCl
phase to an intermediate orthorhombic phase (\textit{Cmcm }or\textit{\ Pbnm}%
),\cite{Luo94,Nelmes95,Mcmahon96} we have examined the XRD pattern obtained at
16 GPa for similar types of distortions with \textit{Cmcm} and \textit{Pbnm}
structures. Both \textit{Pbnm} and \textit{Cmcm} structures were considered
during the Rietveld refinement, and a reliable fit has been obtained using the
\textit{Cmcm} space group, with the Ag/Sb atoms located at 4a sites $(0,0,0)$
and Te atoms occupying the 8d sites $(0.75,0.25,0.5)$. The cell parameters
obtained for the orthorhombic phase are $a=3.7974(3)$ \AA , $b=4.238(4)$
\AA \ and $c=5.6358(1)$ \AA . We find no considerable volume change between
the B1 and orthorhombic phases as the two phases are found to coexist and the
transition is continuous. Our results unambiguously show the distortions are
related to the \textit{Cmcm} space group, and the features of AgSbTe$_{2}$
before amorphization indicate the distortions are similar to those observed in
lead, calcium, and cadmium chalcogenides during the B1-B2
transition.\cite{Luo94,Nelmes95,Mcmahon96,Knorr03} These results closely agree
with the earlier theoretical models explaining the rearrangement of atoms
displaced between the layers during the B1-B2 transitions.%
%TCIMACRO{\FRAME{ftbpFU}{2.3696in}{3.8043in}{0pt}{\Qcb{(a) Sequence of x-ray
%diffraction patterns collected at different pressures while decompression. (b)
%Reappearance of diffraction lines corresponding to the B1 phase after heating
%the DAC for $100^{\circ}$C for 1 hr at ambient pressure. }}{\Qlb{decomp}%
%}{kumar2.eps}{\special{ language "Scientific Word";  type "GRAPHIC";
%maintain-aspect-ratio TRUE;  display "USEDEF";  valid_file "F";
%width 2.3696in;  height 3.8043in;  depth 0pt;  original-width 2.3289in;
%original-height 3.7559in;  cropleft "0";  croptop "1";  cropright "1";
%cropbottom "0";  filename 'kumar2.eps';file-properties "XNPEU";}}}%
%BeginExpansion
\begin{figure}
[ptb]
\begin{center}
\includegraphics[
height=3.8043in,
width=2.3696in
]%
{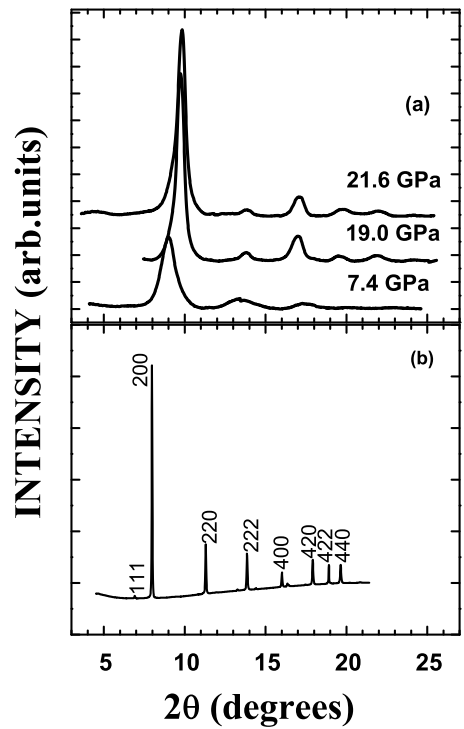}%
\caption{(a) Sequence of x-ray diffraction patterns collected at different
pressures while decompression. (b) Reappearance of diffraction lines
corresponding to the B1 phase after heating the DAC for $100^{\circ}$C for 1
hr at ambient pressure. }%
\label{decomp}%
\end{center}
\end{figure}
%EndExpansion

While comparing our results with similar compounds exhibiting pressure induced
amorphization, it is interesting to note that the corner linked tetrahedral
$\alpha$-quartz and isostructural compounds undergo transitions to a
metastable polymorphic state before amorphization similar to AgSbTe$_{2}$. In
their molecular dynamics simulations based on enthalpy considerations, Chaplot
and Sikka,\cite{Chaplot93} and later, Nandhini Garg and Surinder
Sharma,\cite{Garg00} showed that above 12 GPa, the \textit{Cmcm} phase is
favorable. The experimental observation of the disordered \textit{Cmcm} phase
before amorphization in $\alpha$-FePO$_{4}$ by Pasternak \textit{et
al}.,\cite{Pasternak97} and in GaPO$_{4}$ by other groups, confirm this
scenario.\cite{Badro97} In conjunction with the above, the orthorhombic
\textit{Cmcm} phase is energetically favoured in AgSbTe$_{2}$ while
approaching from B1 to a denser phase. Since simultaneous occurrence of
crystalline and amorphous phases has been seen in FePO$_{4}$, and also partial
amorphization is reported in Co(OH)$_{2}$ by Nguyen \textit{et al}%
.,\cite{Nguyen97} we have examined the relative abundance of the crystalline
phase if any, in the amorphous region. As seen in Fig. 1(c), there are no
significant crystalline features\ identified, only a major halo peak. This
fact eliminates the possibility of coexisting crystalline and amorphous
phases.\ Interestingly, at 26 GPa, the diffraction lines started to reappear,
showing the transformation of the amorphous state to a new crystalline phase.
The observed x-ray diffraction pattern showed few peaks, and the refinement
clearly indicated the existence of a CsCl phase with cell parameter
$a=3.4500(3)$ \AA . The CsCl phase was found to be stable up to 30 GPa as
shown in the Fig. 1(d).

\ The observation of \textit{Cmcm} orthorhombic distortions and an
intermediate amorphous phase, while undergoing transition from the six fold
coordinated B1 phase to eight fold coordinated B2 phase, clearly indicates a
coordination frustration taking place in AgSbTe$_{2}$ due to the defect cubic
structure and movement of \ Te atoms under pressure. This fact is consistent
with the temperature driven amorphous state, observed in the doped family,
where doping may increase internal defects, so that amorphization can be
easily induced by temperature. Even though the structural transitions\ in
AgSbTe$_{2}$ resemble SnI$_{4}$,\cite{Hamaya97} the mechanism of amorphization
is rather different. In the latter compound, molecular dissociation similar to
BaAs is observed where a kinetically frustrated solid state amorphization is
due to a change from fourfold to sixfold coordination.\cite{Greene94}

Selected diffraction patterns during decompression \ are shown in the upper
panel of Fig. 2. The crystalline peaks corresponding to the CsCl phase
remained down to 12 GPa. Below 12 GPa, we observed broadening of the peaks,
and transformation to amorphous state, showing hysteresis in the transition
from CsCl phase to amorphous phase. The amorphous features remained even after
releasing the pressure in the DAC, with a partial crystallization to B1 phase.
The DAC was heated in an oven for 100$^{\circ}$C for 1 hr, and diffraction
patterns were taken. Surprisingly, sharp diffraction lines corresponding to B1
phase reappeared as shown in the bottom panel of Fig. 2, and the B1 phase is
retained with a lattice constant consistent with the initial ambient pressure
value. These facts suggest that the pressure induced amorphous state in
AgSbTe$_{2}$ is reversible, and thermal annealing supplies the energy to
overcome the amorphous-crystalline energy barrier. This activates the
recrystallization process back to the B1 phase identical to that observed in
AlPO$_{4}$ and KH$_{2}$PO$_{4}$ (KDP).\cite{Kruger90,Kobayashi02}

In order to understand the stability of B1 and B2 phases observed
experimentally, we have performed density functional theoretical (DFT)
calculations using the FHI98MD code,\cite{Bockstedte97,Fuchs99} local density
approximation (LDA),\cite{Perdew81} and norm conserving pseudo potentials.
Troullier-Martins type pseudopotentials were generated for Ag, Sb, and Te
atoms and extensive convergence tests were carried out in order to check the
dependency of the number of $k$-points and plane waves.\cite{Troullier91} The
electronic wave-functions were expanded in plane-wave basis set with an energy
cutoff of 71 Ry. Energy convergence was achieved with an accuracy of 10$^{-4}$
eV. The $k$-point integration was performed with mesh points corresponding to
8 $k$-points in the Brillouin zone.%

%TCIMACRO{\FRAME{ftbpFU}{3.0286in}{2.9438in}{0pt}{\Qcb{$P-V$ data for
%AgSbTe$_{2}$. The open symbols (circles denote B1 and diamonds B2) are the
%experimental data and the solid symbols represent the theoretical calculations
%for both phases. The amorphous phase is observed in the shaded region. (Inset)
%Binding energy curves for the B1 and B2 phases of AgSbTe$_{2}$.}}{\Qlb{pv}%
%}{kumar3.eps}{\special{ language "Scientific Word";  type "GRAPHIC";
%maintain-aspect-ratio TRUE;  display "USEDEF";  valid_file "F";
%width 3.0286in;  height 2.9438in;  depth 0pt;  original-width 2.9853in;
%original-height 2.9006in;  cropleft "0";  croptop "1";  cropright "1";
%cropbottom "0";  filename 'kumar3.eps';file-properties "XNPEU";}}}%
%BeginExpansion
\begin{figure}
[ptb]
\begin{center}
\includegraphics[
height=2.9438in,
width=3.0286in
]%
{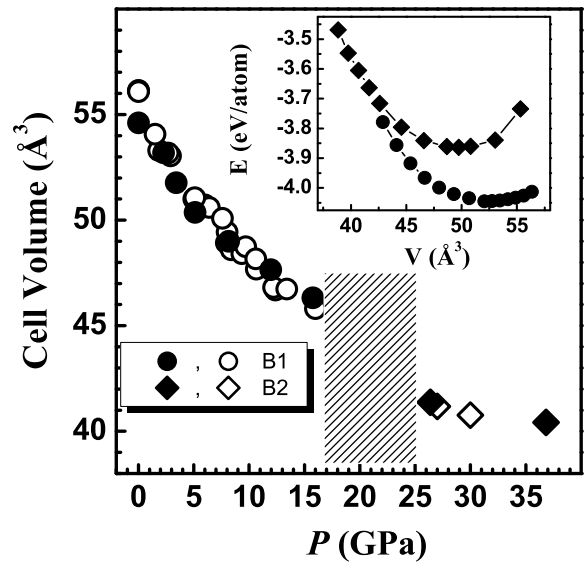}%
\caption{$P-V$ data for AgSbTe$_{2}$. The open symbols (circles denote B1 and
diamonds B2) are the experimental data and the solid symbols represent the
theoretical calculations for both phases. The amorphous phase is observed in
the shaded region. (Inset) Binding energy curves for the B1 and B2 phases of
AgSbTe$_{2}$.}%
\label{pv}%
\end{center}
\end{figure}
%EndExpansion
Binding energy curves show that the B1 phase is energetically favorable by
0.18 eV/atom relative to the B2 phase at ambient conditions as shown in Fig.
3. \ The calculated lattice constant of the B1 phase is 5.93 \AA . This is in
excellent agreement with the experimental value of 6.08 \AA \ and compares
favorably to other calculated values of 5.676 \AA \ and 6.29 \AA \ by M. Luo
\textit{et al}.\cite{Luo04} and R. Detemple \textit{et al}.,\cite{Detemple03}
respectively. Calculations performed for the B2 phase reveal that it is
favoured around 26 GPa. More detailed calculations to understand the
intermediate phases are in progress.

A fit of the experimental $P-V$\ data for the B1 phase, with the second order
Birch-Murnaghan equation of state, yielded a bulk modulus of $B_{0}=45(2)$ GPa
with $B_{0}^{\prime}=4.8$. Even though the lattice constant is underestimated
2\% compared to the experimental value, the bulk modulus 44.5 GPa, obtained
theoretically, is in good agreement with the experimental value, when compared
with the previous reports. \ For direct comparison of the equation of state,
the theoretical volume is scaled to the experimental values as shown in Fig. 3.

In conclusion, we have demonstrated a new structural sequence in the
functionally graded material AgSbTe$_{2}$, which exhibits a B1 to B2
transition with an intermediate amorphous phase. The results of the ab initio
simulations strongly support the pressure induced structural transitions.

The authors thank Dr. Maddury Somayazulu for his technical help at HPCAT. Work
at UNLV is supported by DOE - EPSCoR-State/National Laboratory Partnership
Award DE-FG02-00ER45835 and DOE, NNSA, under Cooperative Agreement
DE-FC08-01NV14049. HPCAT is collaboration among the UNLV High Pressure Science
and Engineering Center, Lawrence Livermore National Laboratory, the
Geophysical Laboratory of the Carnegie Institution of Washington, and the
University of Hawaii at Manoa. Use of the Advanced Photon Source was supported
by the U.S. Department of Energy, Office of Science, Office of Basic Energy
Sciences, under Contract No.W-31-109-Eng-38. This work was also partially
funded by the National Science Foundation (DMR-0072067) and the Office of
Naval Research (N00421-03-0085).

\bibliographystyle{apsrev}
\bibliography{agsbte2,cornel}

\end{document}